\documentclass[11pt]{article}

\usepackage[margin=1in]{geometry}
\usepackage{amsmath,amsfonts,amsthm}
\usepackage{enumitem}
\usepackage{xspace}
\usepackage{url}
\usepackage{hyperref}
\usepackage{amsrefs}
\renewcommand{\eprint}[1]{\href{http://arxiv.org/abs/#1}{#1}}

\def\ba#1\ea{\begin{align}#1\end{align}}

\newtheorem{theorem}{Theorem}
\newtheorem{lemma}[theorem]{Lemma}
\theoremstyle{definition}

\newcommand{\eq}[1]{(\ref{eq:#1})}
\renewcommand{\sec}[1]{Section~\ref{sec:#1}}
\newcommand{\thm}[1]{Theorem~\ref{thm:#1}}
\newcommand{\lem}[1]{Lemma~\ref{lem:#1}}

\newcommand{\gate}[1]{\textsc{#1}\xspace}
\newcommand{\nand}{\gate{nand}}

\newcommand{\E}{\mathbb{E}}
\DeclareMathOperator{\ADV}{ADV}
\newcommand{\norm}[1]{\|{#1}\|}
\newcommand{\floor}[1]{\lfloor{#1}\rfloor}

\begin{document}
\title{The quantum query complexity of certification}
\author{
  Andris Ambainis\thanks{Department of Computing, University of Latvia}
  \and
  Andrew M.\ Childs\thanks{Department of Combinatorics \& Optimization
  and Institute for Quantum Computing, University of Waterloo}
  \and
  Fran\c{c}ois Le Gall\thanks{Quantum Computation and Information Project,
  ERATO-SORST, JST}
  \and
  Seiichiro Tani$^{\ddagger,}$\thanks{NTT Communication Science Laboratories, NTT Corporation}}
\date{}
\maketitle

\begin{abstract}
We study the quantum query complexity of finding a certificate for a $d$-regular, $k$-level balanced \nand formula.  We show that the query complexity is $\tilde\Theta(d^{(k+1)/2})$ for $0$-certificates, and $\tilde\Theta(d^{k/2})$ for $1$-certificates.  In particular, this shows that the zero-error quantum query complexity of evaluating such formulas is $\tilde O(d^{(k+1)/2})$.  Our lower bound relies on the fact that the quantum adversary method obeys a direct sum theorem.
\end{abstract}

\section{Introduction}

Recently, there has been considerable progress in understanding the quantum query complexity of evaluating Boolean formulas.  Here, we consider the closely related problem of \emph{certifying} the value of a formula.

An $n$-bit Boolean formula $\phi$ corresponds to a rooted tree with $n$ leaves, where each leaf represents a binary variable $x_i$ for $i \in \{1,\ldots,n\}$, and each internal vertex represents a logic gate acting on its children.  We can evaluate $\phi$ on an input $x \in \{0,1\}^n$ by applying the logic gates from the leaves toward the root, giving some binary value $\phi(x)$ at the root.  Suppose we are given a black box that, on input $i$, produces the bit $x_i$.  The minimum number of queries to such a black box required to learn $\phi(x)$ for an arbitrary $x \in \{0,1\}^n$ is called the query complexity of evaluating $\phi$.

In this article, we focus on the case of \nand formulas, in which each internal vertex corresponds to a \nand gate.  (Note that formulas consisting of \gate{and}, \gate{or}, and \gate{not} gates are equivalent to \nand formulas.)  Furthermore, we suppose that the tree is regular, with internal vertices of degree $d$, and balanced, with a total of $k$ levels.  Thus, there are $n=d^k$ input variables.

To evaluate such a formula on a black-box input, a deterministic classical computer must make all $n$ queries.  However, the expected number of queries for a randomized classical computer to evaluate the formula with zero error is $O(\lambda_d^k)$ \cite{Sni85,SW86}, where $\lambda_d:=(d-1+\sqrt{d^2+14d+1})/4$, using a simple recursive strategy.  Indeed, this is the optimal query complexity not just for zero-error algorithms \cite{SW86}, but also for bounded-error ones \cite{San95}.  If we fix $d$ and let $k$ grow, this represents a polynomial speedup over the deterministic strategy of evaluating all leaves, since only $O(n^{\log_d \lambda_d})$ queries are required; for example, with $d=2$, the classical query complexity is $\Theta(n^{0.753\ldots})$.  On the other hand, if we fix $k$ and let $d$ grow, then the query complexity is essentially $n$.

For quantum computers, the situation is quite different.  Grover's algorithm shows that if $k=1$, so that $d=n$, a quantum computer can evaluate the formula with bounded error in $O(\sqrt{n})$ queries \cite{Gro97}, which is optimal \cite{BBBV97}.  Similar results can be obtained for constant $k$: a recursive strategy with straightforward error reduction uses $O(\sqrt{n} \log^{k-1}n)$ queries \cite{BCW98}, and a more delicate approach to error reduction uses $O(\sqrt{n} \alpha^k)$ queries for some constant $\alpha$ \cite{HMW03}.  In particular, this shows that the quantum query complexity is $O(\sqrt{n})$ provided $k$ is constant, and even gives some speedup for sufficiently large constant $d$.  However, for the case where $d$ is a small constant, no better-than-classical quantum algorithm was known for nearly ten years.  A breakthrough occurred with the revolutionary quantum walk approach of \cite{FGG07}, which---combined with a simple observation on the simulation of Hamiltonian dynamics by quantum circuits \cite{CCJY07}---showed that $\sqrt{n^{1+o(1)}}$ queries suffice for the case where $d=2$.  Subsequent work showed that in fact the quantum query complexity is $O(\sqrt{n})$ for any $d,k$ \cite{ACRSZ07}.  Since this is optimal \cite{BS04}, the quantum query complexity of evaluating balanced \nand formulas is now a closed problem.

Here, we consider the problem of determining a minimal certificate for the value of a \nand formula.  A \emph{certificate} for $\phi$ on input $x \in \{0,1\}^n$ is a subset $c \subseteq \{1,\ldots,n\}$ such that the values of $x_i$ for $i \in c$ determine $\phi(x)$.  We say that a certificate $c$ is \emph{minimal} if no proper subset of $c$ is also a certificate.

We can view the size of the minimal certificate as a kind of nondeterministic query complexity, since this is the number of queries required to verify the value of the formula if we can guess the certificate.  The certificate size of a regular, balanced \nand formula can be understood simply, as follows.
Let $C_b(k)$ denote the size of a minimal certificate for a $k$-level balanced \nand formula evaluating to $b$ (a \emph{$b$-certificate}).  We have $C_0(0)=C_1(0)=1$, and
\ba
  C_0(k)&=d \, C_1(k-1) &
  C_1(k)&=C_0(k-1).
\ea
Solving this recursion, we obtain
\ba
  C_0(k)&=\begin{cases}d^{k/2} & \text{$k$ even} \\
                       d^{(k+1)/2} & \text{$k$ odd}\end{cases} &
  C_1(k)&=\begin{cases}d^{k/2} & \text{$k$ even} \\
                       d^{(k-1)/2} & \text{$k$ odd}.\end{cases}
\ea
Thus, the certificate size is precisely $\sqrt{n}$ for a formula with an even number of levels, and is close to that value when the number of levels is large and odd.
Note that a minimal $0$-certificate is a set of $0$-inputs for $k$ even and a set of $1$-inputs for $k$ odd; a minimal $1$-certificate is a set of $1$-inputs for $k$ even and a set of $0$-inputs for $k$ odd.

Given a regular, balanced $\nand$ formula and a black box for the input, we are interested in the number of queries required to output a minimal certificate.
This question is of interest for several reasons.  Whereas the optimal classical algorithm for evaluating a \nand formula produces a certificate in the course of its evaluation, the same does not hold for the known optimal quantum algorithms, so it is natural to ask whether quantum computers have a comparable advantage for certification as they do for evaluation.  (The algorithm of \cite{ACRSZ07} produces a quantum state that is related to certificates, which suggests a potential approach to certificate finding, but it turns out that a simpler strategy is nearly optimal.)  In addition, the query complexity of certificate finding gives an upper bound on the zero-error query complexity of the problem (as emphasized in \cite{BCWZ99}), although it is not known if this bound is optimal in general.

Note that Aaronson introduced a notion of {\em quantum certificate complexity} \cite{Aar08}, which is also concerned with quantum algorithms for certifying the value of a function. However, our setting is different: we study quantum algorithms which output classical certificates, while in Aaronson's work, the quantum algorithm itself is a certificate and no classically-verifiable certificate is provided.

For a constant number of levels $k$, the previous best quantum algorithm for producing a certificate for a \nand formula uses $O(d^{(k+2)/2})$ queries in the case of $0$-certificates, and $O(d^{(k+1)/2})$ in the case of $1$-certificates \cite[Lemma 1]{BCWZ99}.  We improve upon this, and also give a result for non-constant $k$, as follows:

\begin{theorem}\label{thm:mainresult}
  Fix $k>1$ and consider a $d$-regular, $k$-level balanced \nand formula $\phi$ with black-box input $x$.
  The bounded-error and zero-error quantum query complexities of certifying $(\phi,x)$ are
\begin{alignat}{2}
  O(k^2 d^{(k+1)/2}) &\quad\text{and}\quad \Omega(d^{(k+1)/2}) &\quad&\text{if~} \phi(x)=0 \\
  O(k^2 d^{k/2}) &\quad\text{and}\quad \Omega(d^{k/2}) &&\text{if~} \phi(x)=1.
\end{alignat}
In particular, the zero-error quantum query complexity of evaluating $\phi(x)$ is $O(k^2 d^{(k+1)/2})$.
\end{theorem}

Note that for $k=1$, the formula is simply a single \nand gate, and the bounded-error query complexity of certifying it on any input is $\Theta(\sqrt n)$.

The upper bound in \thm{mainresult} comes from a strategy that applies the formula evaluation algorithm of \cite{ACRSZ07} recursively, as described in \sec{algorithm}.  The lower bound appeals to direct sum theorems.  We discuss a direct sum theorem for adversary lower bounds in general in \sec{advdirect}, and apply it to certification in \sec{lowerbound}.  We conclude in \sec{open} by mentioning a few open questions.

\section{Algorithm}
\label{sec:algorithm}

To find a certificate for a \nand formula, consider the following simple recursive strategy.  First evaluate the root using the $O(\sqrt n)$-query algorithm of \cite{ACRSZ07}.  If the result is $0$, each child of the root must evaluate to $1$, and we can find a $1$-certificate for each in turn.  If the result is $1$, then we search for a child evaluating to $0$ using Grover's algorithm combined with formula evaluation, and then find a $0$-certificate for it.

This strategy gives a nearly optimal algorithm for certificate finding.  However, we must be careful when applying the formula evaluation algorithm recursively, since it only succeeds with bounded error, and we must ensure that this error can be kept under control.

To find a $b$-certificate for a $k$-level tree, we can apply the procedure $A_b(k)$, defined as follows.

\medskip

\noindent
\textbf{Procedure \boldmath${A_0(k)}$.}
\begin{itemize}[topsep=.5ex,itemsep=0pt]
\item If $k=1$, query every leaf.  If they are all $1$, then output every index; otherwise, go into an infinite loop.
\item If $k>1$, run $A_1(k-1)$ on each subtree.
\end{itemize}

\medskip

\noindent
\textbf{Procedure \boldmath${A_1(k)}$.}
\begin{itemize}[topsep=.5ex,itemsep=0pt]
\item If $k=1$, repeat Grover's algorithm until a $0$-leaf is found, verifying the result to ensure one-sided error.  When a $0$-leaf is found, output its index.
\item If $k>1$, repeat the following until passing the verification:
  \begin{itemize}[topsep=0pt,itemsep=0pt]
  \item Perform robust quantum search~\cite{HMW03} for a 0-subtree, using the \nand tree evaluation algorithm \cite{ACRSZ07} as a subroutine.  (Note that robust quantum search has two-sided error.)  Repeat the search until a $0$-subtree $T$ is found.
\item Verify that $T$ is a 0-subtree with error probability at most $1/n^2$ by running the \nand tree evaluation algorithm $O(\log n)$ times and taking a majority vote.
  \end{itemize}
\item Once $T$ passes the verification, run $A_0(k-1)$ on $T$.
\end{itemize}

\medskip

A simple inductive argument shows that $A_b(k)$ will never terminate when given a formula evaluating to $\bar b$.  Furthermore, when $A_b(k)$ does halt, it is guaranteed to return a correct $b$-certificate.

It remains to understand how long it takes for $A_b(k)$ to produce a $b$-certificate.

\begin{lemma}
\label{lem:querycomplexity}
Suppose that $A_b(k)$ is given a $d$-regular, $k$-level balanced \nand formula evaluating to $b$.  Assume that the verification step in procedure $A_1$ never makes an error.  Then the expected number of queries before $A_0(k),A_1(k)$ terminates is
\ba
O\Big(k\big(1+k\tfrac{\log d}{\sqrt{d}}\big) \sqrt{d^{k+1}}\Big),\quad
O\Big(k\big(1+k\tfrac{\log d}{\sqrt{d}}\big) \sqrt{d^{k}}\Big),
\ea
respectively.
\end{lemma}

\begin{proof}
Let $\E[A_b(k)]$ denote the expected number of queries made by $A_b(k)$ when given a formula evaluating to $b$, assuming the procedure $A_1$ never makes an erroneous verification.  
For $k=1$, we have $\E[A_0(1)]=d$ and $\E[A_1(1)]=O(\sqrt d)$.  For $k \ge 2$,
\begin{align}
  \E[A_1(k)] & = O(d^{k/2}) + O(d^{(k-1)/2} \log n) +  \E[A_0(k-1)] \\
  & = O\Big( \big(1 + k \tfrac{\log d}{\sqrt{d}}\big) d^{k/2} \Big) + \E[A_0(k-1)]
\end{align}
and
\begin{align}
  \E[A_0(k)] = d \, \E[A_1(k-1)].
\end{align}
Solving this recurrence gives
\begin{align}
  \E[A_1(k)] = O\Big(k\big(1+k\tfrac{\log d}{\sqrt{d}}\big)\sqrt{d^k}\Big)
\end{align}
and
\begin{align}
  \E[A_0(k)]
  = O\Big(k\big(1+k\tfrac{\log d}{\sqrt{d}}\big)\sqrt{d^{k+1}}\Big)
\end{align}
as claimed.
\end{proof}

Of course, we cannot actually assume that $A_1$ never makes a verification error, but we can nevertheless obtain a zero-error algorithm for $b$-certification as follows.  First of all, if $A_1$ outputs a certificate, then the certificate is guaranteed to be correct. Thus, a verification error can only result in $A_1$ not finding a certificate (and, possibly, going into an infinite loop while searching for a certificate in the wrong subtree).

To handle the possibility of a verification error causing $A_1$ to run forever,  fix some sufficiently large constant $c$.  Run $A_b$ on the input; if it does not halt after $ck(1+k {\log d}/{\sqrt{d}})d^{(k+1)/2}$ queries (for $b=0$) or $ck(1+k {\log d}/{\sqrt{d}}) d^{k/2}$ queries (for $b=1$), restart $A_1$. The probability that one or more verifications fail during each trial is $O((\log^2 n)/n)$, since the error probability of each verification is at most $1/n^2$, and the total number of verifications per trial is at most the total number of queries, which is upper bounded by $O(n \log^2 n)$.  Thus  only the expected number of $O(1)$ repetitions of $A_1$ are required before the process terminates with a correct certificate. This gives a zero-error quantum algorithm with an expected running time given by \lem{querycomplexity}, with no assumptions about verification errors.

\section{A direct sum theorem for the adversary method}
\label{sec:advdirect}

We now turn to the question of lower bounds.  Our approach to the lower bound for certificate finding is based on the concept of a \emph{direct sum theorem}.  A direct sum theorem says that if solving one instance of a problem requires $q$ queries, then solving $t$ instances requires $\Omega(tq)$ queries.
(Several papers on quantum lower bounds---e.g., \cite{Amb05,KSW07}---have studied direct \emph{product} theorems, which are stronger statements.  In particular, a \emph{strong direct product theorem} says that if an algorithm attempts to solve $t$ instances with $o(tq)$ queries, then it can only succeed with an exponentially small probability.  In this paper, we only need direct sum theorems.)

One of the main methods for proving lower bounds on quantum query complexity is the quantum adversary method \cite{Amb02}, which can be reformulated as follows \cite{BSS03}:

\begin{theorem}
Let $S \subseteq \{0,1\}^n$, and let $\Sigma$ be a finite set.  For any function $f:S \to \Sigma$, define
\ba
  \ADV(f) := \max_{\Gamma} \frac{\norm{\Gamma}}{\max_{i=1}^n \norm{\Gamma \circ D_i}},
\ea
where $\norm{\cdot}$ denotes the spectral norm; $\Gamma \ne 0$ is a symmetric, entrywise non-negative $|S| \times |S|$ matrix satisfying $\Gamma_{xy}=0$ if $f(x)=f(y)$; $(D_i)_{xy} = 1$ if $x_i \ne y_i$ and $(D_i)_{xy} = 0$ otherwise; and $\circ$ denotes the Hadamard (i.e., entrywise) product of matrices.  Finally, let $Q_\epsilon(f)$ denote the minimum number of quantum queries to $x$ needed to compute $f(x)$ with error at most $\epsilon$.  Then
\ba
  Q_\epsilon(f) \ge \frac{1-2\sqrt{\epsilon(1-\epsilon)}}{2} \ADV(f).
\ea
\end{theorem}

We show that the quantity $\ADV(f)$ obeys a direct sum theorem, as follows.

\begin{theorem} \label{thm:advdirect}
  Let $S \subseteq \{0,1\}^n$, and let $\Sigma$ be a finite set.
  Given a function $f^{(1)}:S \to \Sigma$,
  let $f^{(t)}:S^t \to \Sigma^t$ be the function defined by $f^{(t)}(x) := (f^{(1)}(x_1 \ldots x_n),\ldots,f^{(1)}(x_{(t-1)n+1} \ldots x_{tn}))$.  Then
   $\ADV(f^{(t)}) = t\ADV(f^{(1)})$.
\end{theorem}

\begin{proof}
Let $\Gamma^{(1)}$ be an optimal adversary matrix for $f^{(1)}$, and let $\Gamma^{(t)} := \Gamma^{(1)} \otimes I^{\otimes t-1} + I \otimes \Gamma^{(t-1)}$, where $I$ denotes the $|S| \times |S|$ identity matrix.  Choosing $\Gamma=\Gamma^{(t)}$, the numerator of $\ADV(f^{(t)})$ is $\norm{\Gamma^{(t)}}=t\norm{\Gamma^{(1)}}$ (since $\Gamma^{(t)}$ is a sum of $t$ commuting terms, each a tensor product of $\Gamma^{(1)}$ and an identity matrix).

Now let $D_i^{(t)}$ denote the $|S|^t \times |S|^t$ matrix with $(D_i^{(t)})_{xy}=1$ if $x_i \ne y_i$ and $(D_i^{(t)})_{xy}=0$ otherwise.  For $i \in \{1,\ldots,n\}$, we have $D_i^{(t)} = D_i^{(1)} \otimes J^{\otimes t-1}$, where $J$ denotes the $|S| \times |S|$ matrix with every entry equal to $1$.  Matrices $D_i^{(t)}$ for $i \in \{n+1,\ldots,tn\}$ can be expressed similarly by suitable permutation of the tensor factors.  Then for $i \in \{1,\ldots,n\}$,
\ba
  \norm{\Gamma^{(t)} \circ D_i^{(t)}}
  &= \norm{(\Gamma^{(1)} \otimes I^{\otimes t-1} + I \otimes \Gamma^{(t-1)})
      \circ(D_i^{(1)} \otimes J^{\otimes t-1})} \\
  &= \norm{(\Gamma^{(1)} \circ D_i^{(1)}) \otimes I^{\otimes t-1}} \\
  &= \norm{\Gamma^{(1)} \circ D_i^{(1)}},
\ea
where the second equality uses the fact that $I \circ D_i^{(1)}$ is the zero matrix.  By symmetry, $\norm{\Gamma^{(t)} \circ D_i^{(t)}}$ for general $i \in \{1,\ldots,tn\}$ only depends on the congruence class of $i$ modulo $n$.  It follows that $\ADV(f^{(t)}) \ge t \ADV(f^{(1)})$.

Although we do not need the converse to establish a direct sum theorem, it can be proven as follows.  Using semidefinite programming duality, we can express $\ADV(f)$ as a minimization problem; in particular \cite{SS06},
\ba
  \ADV(f) = \min_p \max_{\substack{x,y \in S \\ f(x) \ne f(y)}} \frac{1}{\displaystyle\sum_{i:x_i \ne y_i} \!\!\sqrt{p_x(i) \, p_y(i)}},
\label{eq:advmin}
\ea
where the minimization is over sets of probability distributions $p_x(i)$, i.e., $\sum_{i=1}^n p_x(i) = 1$ for all $x \in S$.  Let $p^{(1)}$ achieve the minimum in \eq{advmin} for $f=f^{(1)}$.  For each $x \in S^t$, let $p_x^{(t)}$ be the probability distribution obtained by selecting one of the $t$ blocks uniformly at random and then choosing the $i$th coordinate within that block according to $p^{(1)}$.  In other words, let $p^{(t)}_x(i) := \frac{1}{t} p^{(1)}_{x_{n\floor{i/n}+1} \ldots x_{n\floor{i/n}+n}}(i \bmod n)$ for each $i \in \{1,\ldots,tn\}$.  To have $f^{(t)}(x) \ne f^{(t)}(y)$, we must have $f^{(1)}(x_{jn+1} \ldots x_{jn+n}) \ne f^{(1)}(y_{jn+1} \ldots y_{jn+n})$ for some $j \in \{0,\ldots,t-1\}$; without loss of generality, suppose $j=1$.  Since the denominator of \eq{advmin} is a sum of positive terms, the maximum over $x,y \in S^t$ is achieved by the case where $x_i=y_i$ for all $i \in \{n+1,\ldots,tn\}$, such that there is no contribution to the denominator from these indices.  Thus, we find
\ba
  \ADV(f^{(t)})
  &\le \max_{\substack{x,y \in S \\f^{(1)}(x) \ne f^{(1)}(y)}}
       \frac{1}{\displaystyle\sum_{i:x_i \ne y_i} \!\!
       \sqrt{\frac{1}{t}p_x^{(1)}(i) \, \frac{1}{t} p_y^{(1)}(i)}} 
   = t \ADV(f^{(1)}),
\ea
which completes the proof.
\end{proof}

As a consequence of this fact, whenever a function $f$ has an optimal adversary lower bound, it obeys a direct sum theorem.  This can be compared with the multiplicative adversary method \cite{Spa08}, in which a multiplicative adversary lower bound implies a strong direct product theorem.

Note that the lower bound of \thm{advdirect} applies to the negative adversary method as well \cite{HLS07}, so negative adversary lower bounds also imply direct sum theorems.

\section{Lower bound}
\label{sec:lowerbound}

We now apply the result of the previous section to give a nearly optimal lower bound for certificate finding.

The query complexity of certification is clearly $\Omega(d^{k/2})$, as knowing a certificate allows one to evaluate the formula.  (Recall that a minimal certificate for a balanced formula is a subset of inputs all taking the same value; hence we can learn this value by evaluating one input.)  This immediately shows that the algorithm of \sec{algorithm} is close to optimal for $1$-certificate finding.

It remains to prove the lower bound for $0$-certificate finding.  First, suppose that $k$ is even.  In this case, a $0$-certificate consists of one $0$-leaf from each of $d^{k/2}$ bottom subtrees, where each subtree has size $d$.  Suppose we are told which of the bottom subtrees we need to consider; this can only make the problem easier.  Then we must find a $0$-leaf in each of these subtrees.  Intuitively, the best strategy for finding these leaves is to run Grover's algorithm for an input of size $d$ independently on each of the $d^{k/2}$ subtrees.
To make this intuition rigorous, we can apply the direct sum theorem for search, which says we need at least $\Omega(t \sqrt{d})$ queries to solve $t$ independent instances of the search problem on inputs of size $d$.  
While we could prove this using \thm{advdirect}, it is also implied by the strong direct product theorem for search from \cite{Amb05} (which applies even to the problem of finding $t$ marked items with the promise that they exist).
Thus, with $t=d^{k/2}$, we find a lower bound of $\Omega(d^{(k+1)/2})$ queries.  (A similar argument works for $1$-certificate finding with $k$ odd, but we do not need it.)

We can apply a similar approach when $k$ is odd.  Suppose we could prove a direct sum theorem for $1$-certifying two-level \nand formulas---in other words, that finding $1$-certificates to $t$ independent two-level formulas requires $t$ times the number of queries required to certify one such formula.   A $0$-certificate with $k$ odd consists of all $d$ $1$-leaves of each of $d^{(k-1)/2}$ bottom subtrees.  If we are told which subtrees we must consider one level above, then we must evaluate $d^{(k-1)/2}$ two-level trees.  Thus, a direct sum theorem would say that we need $\Omega(d^{(k-1)/2} \cdot d) = \Omega(d^{(k+1)/2})$ queries.

To prove this direct sum theorem, by \thm{advdirect}, it suffices to give an adversary lower bound for the problem of $1$-certifying a two-level \nand formula.  Since we are promised that the formula evaluates to $1$, at least one of its $d$ subtrees evaluates to $0$, i.e., all $d$ inputs to that subtree are $1$.  We further promise that precisely one subtree evaluates to $0$, and that in all other subtrees, $d-1$ of the $d$ inputs evaluate to $1$, with the remaining input evaluating to $0$.  These promises only make the problem easier.  Let $S \subseteq \{0,1\}^{d^2}$ denote the set of such inputs, and let $f:S \to \{1,\ldots,d\}$ be defined by $f(x)=i$, where $i$ is the unique index of the subtree that evaluates to $0$.

We claim that $\ADV(f) = \Omega(d)$.  To see this, let $\Gamma$ be the $|S| \times |S|$ matrix in which $\Gamma_{xy}=0$ unless $x$ and $y$ differ in exactly two bits and $f(x) \ne f(y)$, in which case $\Gamma_{xy}=1$.  In the latter case,
\begin{itemize}[itemsep=0pt]
\item
the subtree which evaluates to $0$ in $x$ evaluates to $1$ in $y$, meaning that one of its $d$ inputs switches from $1$ in $x$ to $0$ in $y$, and
\item
exactly one of $d-1$ subtrees which evaluates to $1$ in $x$ evaluates to $0$ in $y$, meaning that its unique $0$-input in $x$ switches to $1$ in $y$.
\end{itemize}
Thus, for any fixed $x$, there are $d(d-1)$ possible $y$ with $f(x) \ne f(y)$.  In other words, $\Gamma$ is the adjacency matrix of a $d(d-1)$-regular graph, which means that $\norm{\Gamma}=d(d-1)$.

Now, fixing $x$ as well as an index $i$, the maximum number of $y$s with $x_i\neq y_i$ is $d$: if $i$ is an input to the subtree evaluating to $0$ in $x$, then we only have the freedom to choose one of $d-1$ other subtrees in which we switch a fixed input, whereas if $i$ is an input to a subtree evaluating to $1$ in $x$, then we only have the freedom to choose one of $d$ inputs to the $0$-subtree to switch.  This means that $D_i \circ \Gamma$ is the adjacency matrix of a graph of maximum degree $d$, which implies $\norm{D_i \circ \Gamma} \le d$.  Overall, we have $\ADV(f) \ge d-1 = \Omega(d)$.

Note that, more generally, a similar argument can be used to establish a direct sum theorem for the problem of certifying $k$-level \nand formulas.  In particular, the bounded-error quantum query complexity of certifying $t$ independent $d$-regular, $k$-level balanced \nand formulas, each of which evaluates to $b\in \{0,1\}$, is 
$\Omega(td^{(k+1)/2})$ for $b=0$ and
$\Omega(td^{k/2})$ for $b=1$.

\section{Open questions}
\label{sec:open}

We have described a nearly optimal quantum algorithm for certifying the value of a regular, balanced \nand formula.  This leaves several open questions:

\begin{itemize}[itemsep=0pt]
\item For constant-degree formulas, there is a logarithmic gap between our lower and upper bounds.  Is it possible to improve the lower bound or the algorithm (or both)?

\item Since a certificate can be used to verify the value of a formula with zero error in only $O(\sqrt n)$ queries, our result shows in particular that the zero-error quantum query complexity of evaluating a $d$-regular, $k$-level balanced \nand formula is $O(k^2 d^{(k+1)/2})$.  However, the best known lower bound is $\Omega(d^{k/2})$, from the bounded-error case.  Is there a faster quantum algorithm for evaluating a formula with zero error that does not work by producing a certificate?  (More generally, the relationship between bounded- and zero-error quantum query complexity is poorly understood.)

\item We have restricted our attention to regular, balanced formulas.  What is the quantum query complexity of certifying an arbitrary $\nand$ formula?
Of course, we can apply a similar recursive strategy to unbalanced formulas, but in general, it is not clear how Procedure~$A_1$ should select a $0$-subtree, since some subtrees may require more queries to certify than others.
\end{itemize}

\section*{Acknowledgments}

We thank Harry Buhrman and Rahul Jain for helpful discussions.  We also thank Robin Kothari for pointing out that \thm{mainresult} only applies for $k>1$.
AA was supported by University of Latvia research grant ZP01-100 and a Marie Curie International Reintegration Grant (IRG).
AMC received support from MITACS, NSERC, and the US ARO/DTO.



\begin{bibdiv}
\begin{biblist}

\bib{Aar08}{article}{
  author    = {S. Aaronson},
  title     = {Quantum certificate complexity},
  journal   = {Journal of Computer and System Sciences},
  volume    = {74},
  number    = {3},
  year      = {2008},
  pages     = {313-322},
  note      = {Preliminary version in CCC 2003},
  eprint    = {quant-ph/0210020}
}

\bib{Amb02}{article}{
      author={Ambainis, A.},
       title={Quantum lower bounds by quantum arguments},
        date={2002},
     journal={Journal of Computer and System Sciences},
      volume={64},
      number={4},
       pages={750\ndash 767},
        note={Preliminary version in STOC 2000},
      eprint={quant-ph/0002066},
}

\bib{Amb05}{techreport}{
      author={Ambainis, A.},
       title={A new quantum lower bound method, with an application to strong
  direct product theorem for quantum search},
      eprint={quant-ph/0508200},
}

\bib{ACRSZ07}{inproceedings}{
      author={Ambainis, A.},
      author={Childs, A.~M.},
      author={Reichardt, B.~W.},
      author={{\v S}palek, R.},
      author={Zhang, S.},
       title={Any {AND-OR} formula of size {$N$} can be evaluated in time
  {$N^{1/2 + o(1)}$} on a quantum computer},
        date={2007},
   booktitle={Proc. 48th IEEE Symposium on Foundations of Computer Science},
       pages={363\ndash 372},
        note={Merged version of \eprint{quant-ph/0703015} and \eprint{arXiv:0704.3628}},
}

\bib{BS04}{article}{
      author={Barnum, H.},
      author={Saks, M.},
       title={A lower bound on the quantum query complexity of read-once
  functions},
        date={2004},
     journal={Journal of Computer and System Sciences},
      volume={69},
      number={2},
       pages={244\ndash 258},
      eprint={quant-ph/0201007},
}

\bib{BSS03}{inproceedings}{
      author={Barnum, H.},
      author={Saks, M.},
      author={Szegedy, M.},
       title={Quantum query complexity and semidefinite programming},
        date={2003},
   booktitle={Proc. 18th IEEE Conference on Computational Complexity},
       pages={179\ndash 193},
}

\bib{BBBV97}{article}{
      author={Bennett, C.~H.},
      author={Bernstein, E.},
      author={Brassard, G.},
      author={Vazirani, U.},
       title={Strengths and weaknesses of quantum computing},
        date={1997},
     journal={SIAM Journal on Computing},
      volume={26},
       pages={1510\ndash 1523},
      eprint={quant-ph/9701001},
}

\bib{BCW98}{inproceedings}{
      author={Buhrman, H.},
      author={Cleve, R.},
      author={Wigderson, A.},
       title={Quantum vs. classical communication and computation},
        date={1998},
   booktitle={Proc. 30th ACM Symposium on Theory of Computing},
       pages={63\ndash 68},
      eprint={quant-ph/9802040},
}

\bib{BCWZ99}{inproceedings}{
      author={Buhrman, H.},
      author={Cleve, R.},
      author={de Wolf, R.},
      author={Zalka, C.},
       title={Bounds for small-error and zero-error quantum algorithms},
        date={1999},
   booktitle={Proc. 40th IEEE Symposium on Foundations of Computer Science},
       pages={358\ndash 368},
      eprint={cs/9904019},
}

\bib{CCJY07}{techreport}{
      author={Childs, A.~M.},
      author={Cleve, R.},
      author={Jordan, S.~P.},
      author={Yeung, D.},
       title={Discrete-query quantum algorithm for {NAND} trees},
      eprint={quant-ph/0702160},
}

\bib{FGG07}{article}{
      author={Farhi, E.},
      author={Goldstone, J.},
      author={Gutmann, S.},
       title={A quantum algorithm for the {H}amiltonian {NAND} tree},
        date={2008},
     journal={Theory of Computing},
      volume={4},
      number={1},
       pages={169\ndash 190},
      eprint={quant-ph/0702144},
}

\bib{Gro97}{article}{
      author={Grover, L.~K.},
       title={Quantum mechanics helps in searching for a needle in a haystack},
        date={1997},
     journal={Physical Review Letters},
      volume={79},
       pages={325\ndash 328},
      eprint={quant-ph/9706033},
        note={Preliminary version in STOC 1996},
}

\bib{HLS07}{inproceedings}{
      author={H{\o}yer, P.},
      author={Lee, T.},
      author={{\v S}palek, R.},
       title={Negative weights make adversaries stronger},
        date={2007},
   booktitle={Proc. 39th ACM Symposium on Theory of Computing},
       pages={526\ndash 535},
}

\bib{HMW03}{inproceedings}{
      author={H{\o}yer, P.},
      author={Mosca, M.},
      author={de~Wolf, R.},
       title={Quantum search on bounded-error inputs},
        date={2003},
   booktitle={Proc. 30th International Colloquium on Automata, Languages, and
  Programming},
      series={Lecture Notes in Computer Science},
      volume={2719},
       pages={291\ndash 299},
      eprint={quant-ph/0304052},
}

\bib{KSW07}{article}{
      author={Klauck, H.},
      author={{\v S}palek, R.},
      author={de~Wolf, R.},
       title={Quantum and classical strong direct product theorems and optimal
  time-space tradeoffs},
        date={2007},
     journal={SIAM Journal on Computing},
      volume={36},
      number={5},
       pages={1472\ndash 1493},
      eprint={quant-ph/0402123},
        note={Preliminary version in FOCS 2004},
}

\bib{SW86}{inproceedings}{
      author={Saks, M.},
      author={Wigderson, A.},
       title={Probabilistic boolean decision trees and the complexity of
  evaluating game trees},
        date={1986},
   booktitle={Proc. 27th IEEE Symposium on Foundations of Computer Science},
       pages={29\ndash 38},
}

\bib{San95}{article}{
      author={Santha, M.},
       title={On the {M}onte {C}arlo {B}oolean decision tree complexity of
  read-once formulae},
        date={1995},
     journal={Random Structures and Algorithms},
      volume={6},
      number={1},
       pages={75\ndash 87},
}

\bib{Sni85}{article}{
      author={Snir, M.},
       title={Lower bounds on probabilistic linear decision trees},
        date={1985},
     journal={Theoretical Computer Science},
      volume={38},
       pages={69\ndash 82},
}

\bib{Spa08}{inproceedings}{
      author={{\v S}palek, R.},
       title={The multiplicative quantum adversary},
        date={2008},
   booktitle={Proc. 23rd IEEE Conference on Computational Complexity},
       pages={237\ndash 248},
      eprint={quant-ph/0703237},
}

\bib{SS06}{article}{
      author={{\v S}palek, R.},
      author={Szegedy, M.},
       title={All quantum adversary methods are equivalent},
        date={2006},
     journal={Theory of Computing},
      volume={2},
      number={1},
       pages={1\ndash 18},
      eprint={quant-ph/0409116},
        note={Preliminary version in ICALP 2005},
}

\end{biblist}
\end{bibdiv}
\end{document}